\documentclass{appolb}
\usepackage{epsfig}

\begin{document}
\title{Pion and sigma meson dissociation in a modified NJL model at finite 
temperature\thanks{Presented at the "XXXI Max Born Symposium and HIC for FAIR 
Workshop", \\Wroclaw, Poland, June 14-16, 2013 }}
\author{A.~Dubinin
\address{Institute for Theoretical Physics, University of Wroclaw, 
Wroclaw, Poland}
\\[5mm]
D.~Blaschke
\address{
Institute for Theoretical Physics, University of Wroclaw, 
Wroclaw, Poland\\
Bogoliubov Laboratory for Theoretical Physics, JINR Dubna, 
Dubna, Russia}
\\[5mm]
Yu.~L.~Kalinovsky
\address{Laboratory for Information Technologies, JINR Dubna, 
Dubna, Russia}
}
\maketitle
\begin{abstract}
We investigate pion and sigma meson correlations 
in hot quark matter within a modified NJL model. 
Special emphasis is on the transformation of mesonic bound states to resonances
(Mott dissociation) when due to the partial chiral symmetry restoration with 
increasing temperature the threshold of quark-antiquark continuum states drops 
below the meson mass at the corresponding Mott temperature. 
The description is based on evaluating the polarization functions for 
quark-antiquark (meson) correlations as a function of the temperature and
the results can be represented by introducing modulus and phase of the complex 
propagator functions for the mesonic states. 
We study the effect of modelling confinement  by introducing a low-momentum  
cutoff in loop integrals. 
We make the ansatz that this cutoff is identified with the dynamically 
generated quark mass gap and find an increase of the continuum threshold
which makes the otherwise unbound sigma meson a bound state in the vacuum.
We discuss the in-medium behavior of the mesonic phase shifts including the 
Mott effect and find accordance with the Levinson theorem.  
\end{abstract}
  
\section{Introduction}
Experimental data from existing (RHIC Brookhaven, CERN SPS) and planned 
(NICA @ JINR Dubna and CBM @ FAIR Darmstadt) particle accelerators explore the 
characteristics of the hadron to quark  matter phase transition. 
This phase transition is expected to play a crucial role also in the 
astrophysics of compact stars, binary compact star mergers and supernova 
explosions. 
The experimental diagnostics and adequate theoretical description of this 
phenomenon are problems of high actuality.  
The theoretical description of the phase transition region should be based 
on a field theoretical description  of the effective interactions in quark 
matter. 
It is essential for a modern description of quark matter with hadronic bound 
states to implement the features of chiral symmetry breaking, deconfinement 
and color superconductivity.

The Nambu-Jona-Lasinio type model is a field theoretical quark model with 
current-current type interactions adjusted for the description of low energy 
meson and diquark physics
\cite{Nambu:1961tp,Volkov:1984kq,Vogl:1991qt,Klevansky:1992qe,Hatsuda:1994pi,Buballa:2003qv}. 
Within this model the mechanism of spontaneous breaking of chiral symmetry 
(SBCS) is realized in a simple and transparent way, and the low energy 
theorems are fulfilled. 
It is straightforwardly generalized to finite temperatures and chemical 
potentials within the Matsubara formalism which provides results for the 
mean field thermodynamics of quark matter that implements chiral symmetry
restoration. 
The coupling of the chiral quark dynamics to the Polyakov-loop allows to 
suppress the occurrence of free quarks at too low temperatures, before the
chiral symmetry restoration transition.
   
However, the ordinary NJL model as well as its Polyakov-loop counterpart
both fail to prevent low-lying hadron states (like the $\sigma$ or $\rho$ 
meson) from decaying to free quarks, which makes a realistic description of 
hadrons on their mass shell questionable.
In order to cure this problem, we introduce an infrared (IR) cutoff to 
quark momentum integrals since in confined matter the long wavelength 
(low momentum) modes of quark fields shall be absent since quarks are enclosed
in hadrons only. 
For NJL models with IR cutoff see, e.g., 
Refs.~\cite{Ebert:1996vx,Blaschke:1998ws,
GutierrezGuerrero:2010md}.
In such a picture deconfinement occurs when the IR cutoff goes to zero.
As a natural assumption, based on the close relationship of confinement and
chiral symmetry breaking, we will identify the IR cutoff with the dynamical 
quark mass.  

In the present contribution we want to go beyond the mean field level and 
focus on the description of low-lying hadronic bound states such as the 
pion and sigma meson chiral partner system and its dissociation due to the 
Mott effect at finite temperature.
A consistent thermodynamic description of correlations in many-particle 
systems can be achieved with the Beth-Uhlenbeck approach to the virial 
expansion \cite{Beth-Uhlenbeck} and its relativistic formulation 
\cite{Dashen:1969ep}, based on scattering phase shifts. 
The Beth-Uhlenbeck approach has been generalized to address the Mott
dissociation of bound states (see, e.g., \cite{Schmidt:1990}) 
including the case of meson dissociation within the NJL model 
\cite{Hufner:1994ma,Zhuang:1994dw} and its Polyakov-loop generalization
\cite{Rossner:2007ik,Wergieluk:2012gd,Yamazaki:2012ux,Blaschke:2013zaa}

In our contribution to these Proceedings, we will discuss the effect of the
IR cutoff on the quark and meson mass spectrum at finite temperature and 
show that the sigma meson can be obtained as a bound state at low temperatures.
We will evaluate the mesonic scattering phase shifts as basic inputs for the 
generalized Beth-Uhlenbeck equation of state and put special emphasis on the
Mott dissociation effect which is obtained in accordance with the Levinson 
theorem.  
We use here the NJL model and reserve the straightforward coupling to the 
Polyakov loop to a subsequent study.

\section{Mass gap and correlations in a NJL model with IR cutoff  
}

We consider the two-flavor NJL model for quark matter at finite temperature 
$T$ and chemical potential  $\mu$ with the partition function 
\begin{eqnarray}
\label{ZNJL}
Z(T,\mu)= \int {\mathcal D} \bar{q} {\mathcal D} {q}
\exp \bigg\{ \int_0^\beta d\tau \int_V  d^3 x  
&&\Big\{\bar{q} (i\gamma^\mu \partial_\mu-m_{0} - \gamma^0\mu)q
\nonumber \\
&& + G_{S} \left[(\bar{q}q)^2+(\bar{q}i\gamma_{5}\vec{\tau}q)^2
   \right]\Big\}\bigg\}~.
\end{eqnarray}   
Here $q$ and $\bar{q}$ denote the quark spinor fields with antiperiodic 
boundary conditions in the imaginary time interval $0\le \tau\le \beta=1/T$, 
$G_S$  is the coupling constant,
$\vec{\tau}$ is the vector of Pauli matrices in flavor space, 
and $m_0=\mbox{diag}(m_u^0,m_d^0)$ is the diagonal matrix 
of current quark masses. 

The thermodynamic potential $\Omega(T,\mu)=-T \ln Z(T,\mu) / V$ 
in Gaussian approximation is a sum of mean field ({\rm MF}) and 
fluctuation part
\begin{eqnarray}
\Omega(T,\mu)=\Omega_{\rm MF}(T,\mu)+\sum_{M} \Omega^{(2)}_{M}(T,\mu)+\mathcal{O}[\phi^3_M]~,
\nonumber
\end{eqnarray}
with the  mean field part given by
\begin{eqnarray}
&&\Omega_{\rm MF}(T,\mu)
= \frac{\sigma^2_{\rm MF}}{4G_S}
+2N_cN_f\int\frac{d^3p}{(2\pi)^3}
\left[E_p - T\ln f^{+}(E_p) - T\ln f^{-}(E_p) \right]~,
\nonumber
\end{eqnarray}
where $f^{\mp}(E)=[e^{\beta(E\mp\mu)}+1]^{-1} $ 
is the distribution function for fermions (antifermions) with the
dispersion relation $E_p=\sqrt{p^2+m^2}$.

\begin{figure}[h!]
\centerline{
\includegraphics[width=0.7\textwidth,angle=0]{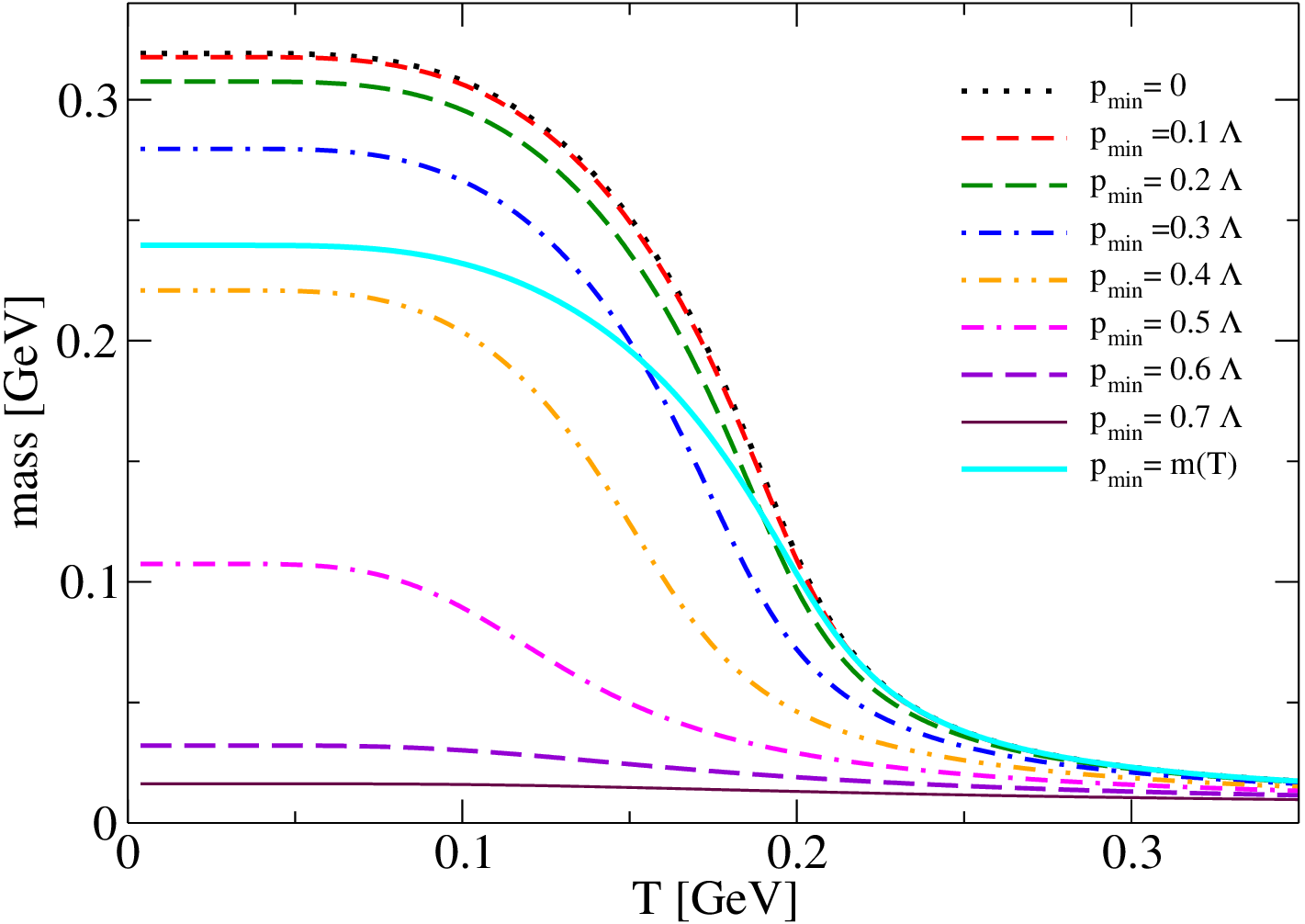}}
\caption{Quark mass vs. temperature for different 
$p_{\rm min} = {\rm const}$ and $p_{\rm min} = m(T)$. 
\label{quarkmass}
}
\end{figure}

The  minimization of mean field part of the  thermodynamic potential,
$\partial\Omega_{\rm MF}/\partial \sigma_{\rm MF}=0$,
 leads to the gap equation ($m=m_0+\sigma_{\rm MF}$)
\begin{eqnarray}
\label{gap}
m=m_0+4G_SN_fN_c\int_{p_{\rm min}}^\Lambda
\frac{dp~p^2}{2\pi^2}\frac{m}{E_p}\Big[1- f^+(E_p)-f^-(E_p) \Big]~.
\end{eqnarray}
Temperature dependent solutions of Eq.~(\ref{gap}) are shown in 
Fig.~\ref{quarkmass} for different choices of the IR cutoff $p_{\rm min}$
which mimicks confinement.

\begin{figure}[h!]
\label{mass}
\centerline{
\includegraphics[width=0.5\textwidth,angle=0]{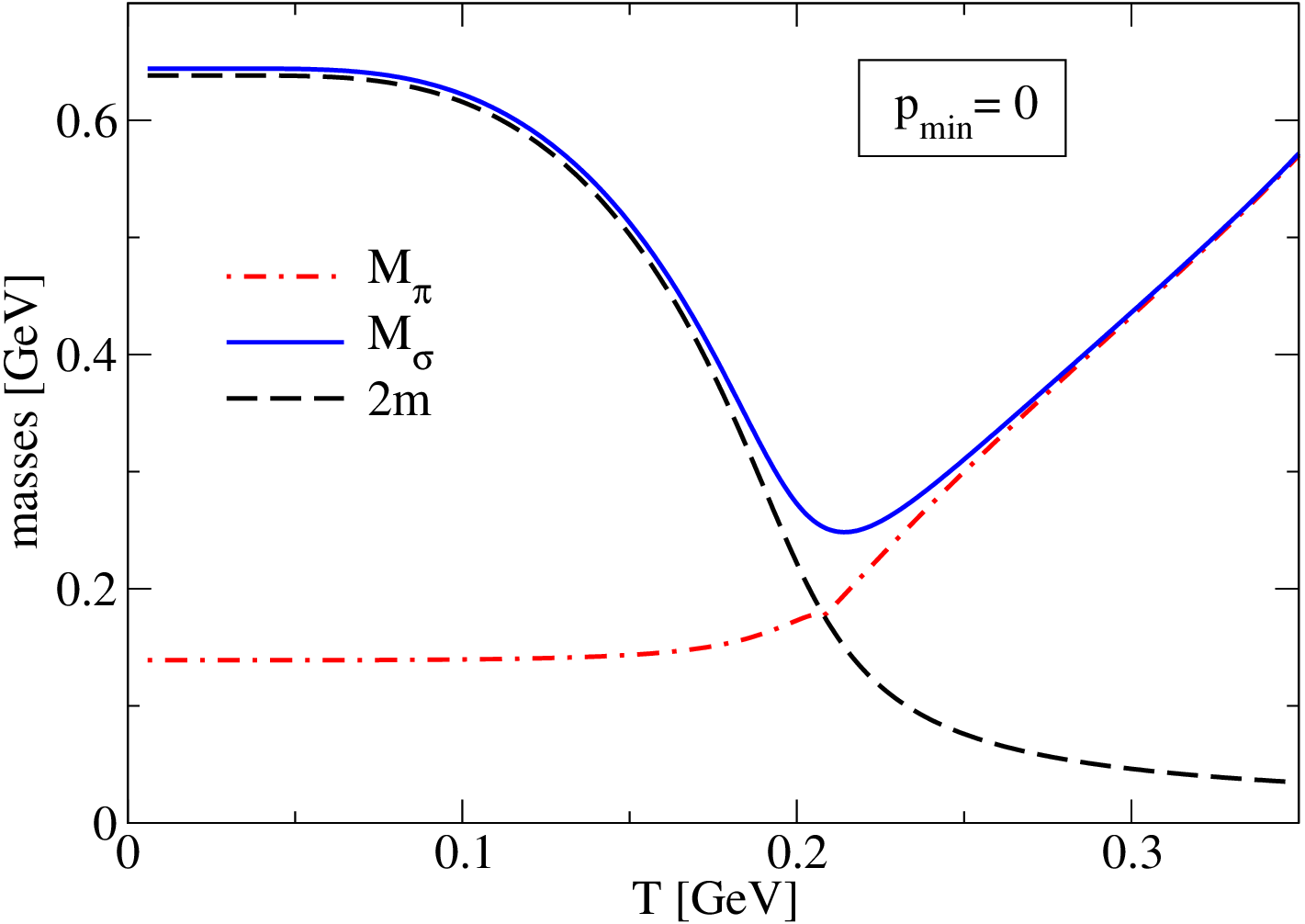}
\includegraphics[width=0.5\textwidth,angle=0]{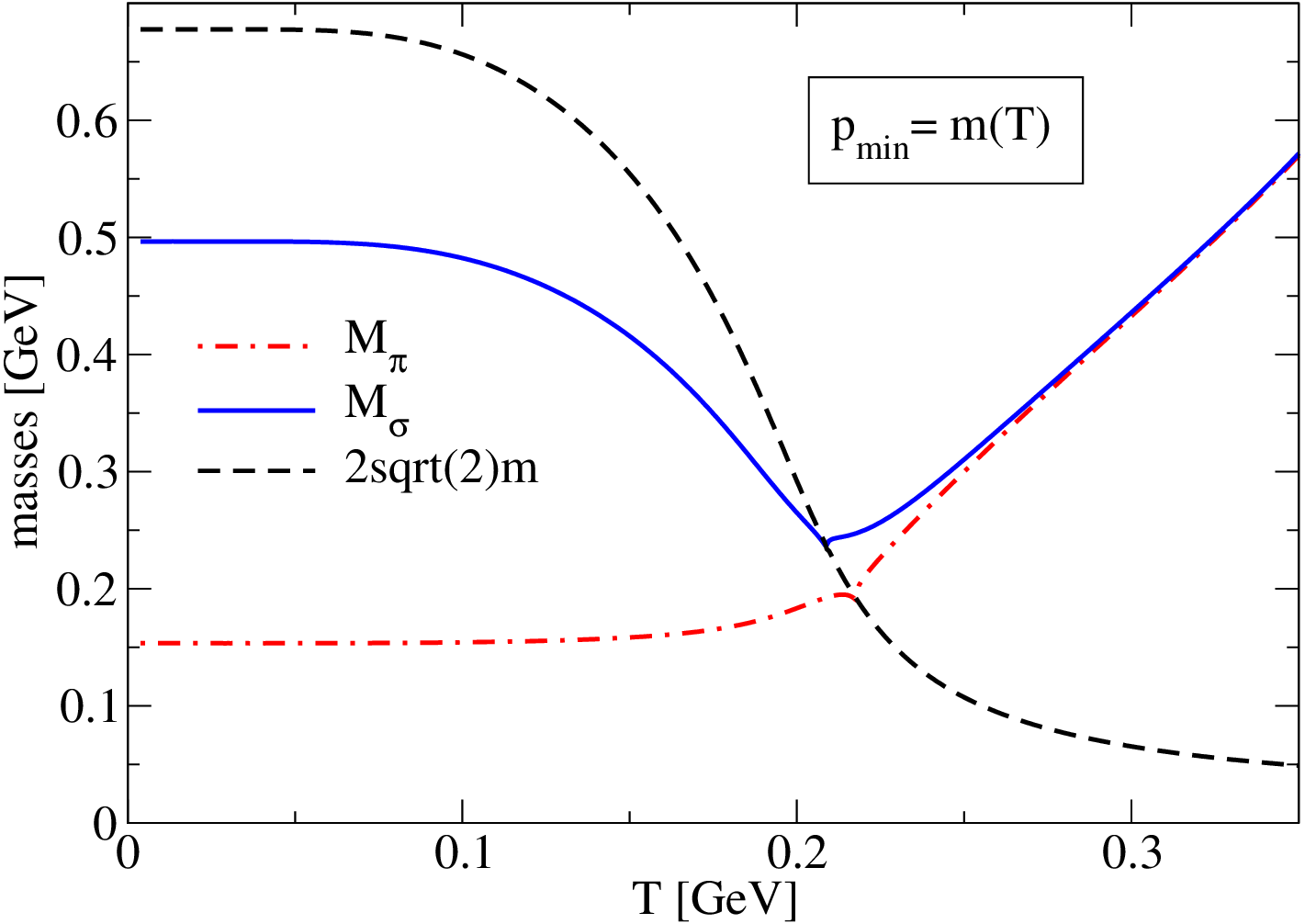}}
\caption{Temperature dependence of the masses for the pion 
(dash-dotted line), sigma meson (solid line) and the continuum 
threshold (dashed line) for $p_{\rm min}=0$ (left panel) and 
for $p_{min}=m(T)$ (right panel).}
\end{figure}

\begin{figure}[t!hb]
\centerline{
\includegraphics[width=0.9\textwidth,angle=0]{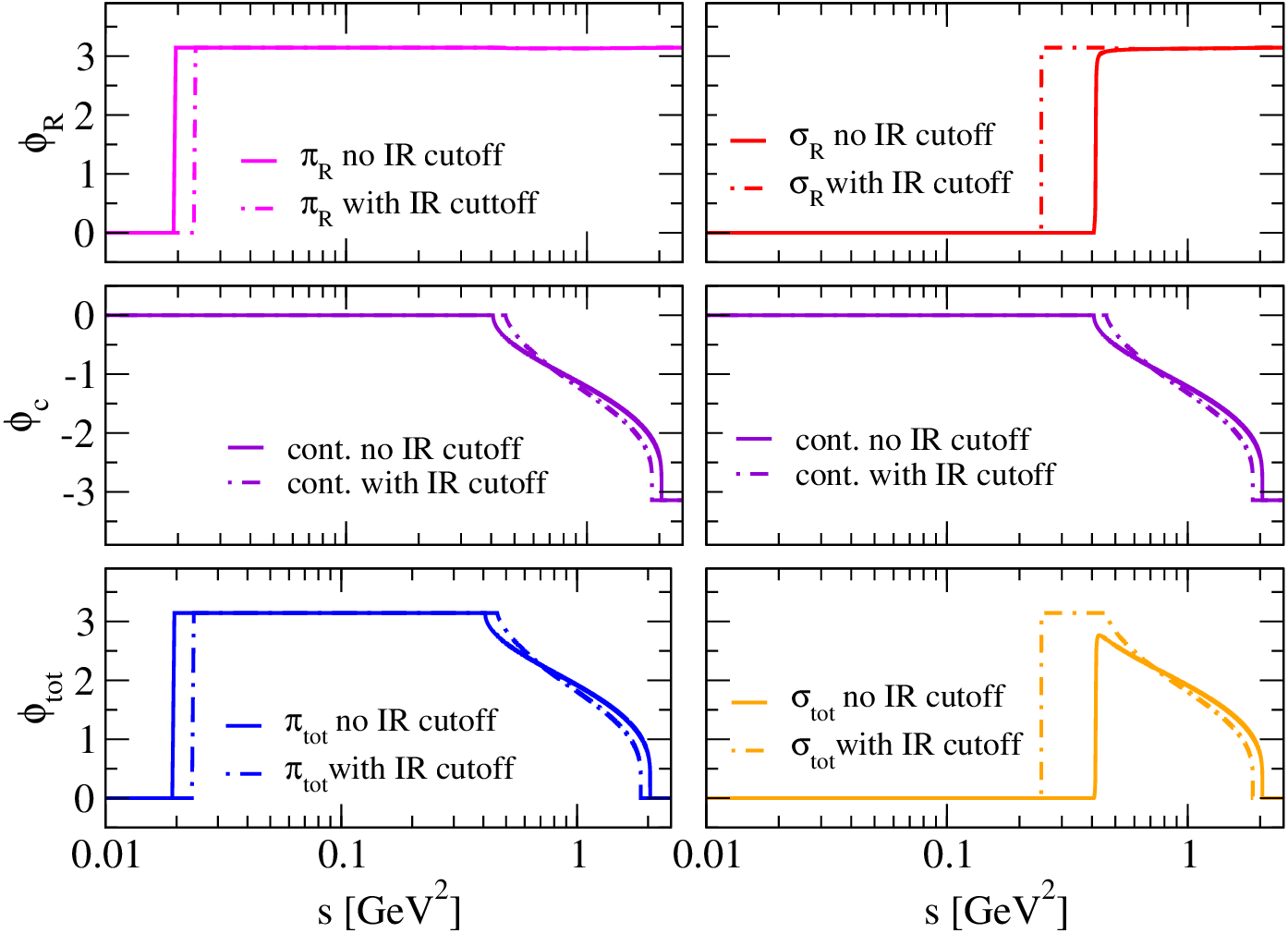}}
\caption{Resonance, continuum and total phase shifts at $T=0$ 
vs. squared center of mass energy $s$
for the pion (left panels) and the sigma meson (right panels) 
with IR cutoff $p_{\rm min}=m(T)$ (dash-dotted lines) and without 
it (solid lines).
\label{phase0}
}
\end{figure}

The fluctuations contribution to the Gaussian order of the
thermodynamic potential  is by definition given by

\begin{eqnarray}
\Omega^{(2)}_{M}(T,\mu)=\sum_M
\frac{N_M}{2}\frac{T}{V}\Tr \ln D^{-1}_M(\omega_n,{q}),
\end{eqnarray}
where $M=\pi,\sigma$, $N_\pi=3$,  $N_\sigma=1$ and the  
inverse meson propagator
$D_M^{-1}(\omega_n,{q})=1/G_S-\Pi_M(\omega_n,{q})$ 
is defined via the polarization loop 
\begin{eqnarray}
\label{pol}
\Pi_{M}(\omega_n,q)&=&- 2N_fN_c\sum_{s_{},s{'}=\pm 1}
\int\frac{d^3p}{(2\pi)^3}
\frac{1-f^{+}(s^{}E_p)-f^{-}(-s^{'}E_k)}
{\omega_n+s^{'}E_{k}-sE_{p}}\chi^{\pm}_{-},
\nonumber\\
\chi^{\pm}_{-} &=& 1-ss{'}\frac{pk\mp m^2}{E_p E_k}~.
\end{eqnarray}
The expressions for the polarization loop contain an integral 
over 3-momentum $p$ and the abbreviation $k=p-q$.
For correlations at rest, $q=0$, we have 
$p=k$ and  obtain the homogeneous Bethe-Salpeter
equations for pion and sigma meson bound states 
($P_\pi=p^2+m^2$, $P_\sigma=p^2$)
from the pole approximation to the analytically 
continued ($i\omega_n\to z$) meson 
propagator $D_M(s)=(s-M^2_M)^{-1}$, 
$s=\omega^2-q^2$, $\omega={\rm Re}(z)$
\begin{eqnarray}
1+4G_SN_cN_f \int_{p_{\rm min}}^\Lambda 
\frac{dp~p^2}{2\pi^2}\frac{1}{E_p}\frac{P_M}{M^2_M-4E^2_p}
\left[1-f^{+}(E_p)-f^{-}(E_p)\right]=0~.
\end{eqnarray} 
The results for the temperature dependence of the meson 
mass spectrum and the continuum threshold are shown in 
Fig.~\ref{mass} for both cases, the ordinary and the modified
NJL model with $p_{\rm min}=0$ and  $p_{\rm min}=m(T)$,
respectively.
Comparing both cases one observes that in the latter the 
sigma meson is a bound state for temperatures below the 
Mott temperature, implicitely defined as 
$M_M(T_{\rm Mott,M})=\sqrt{s_{\rm thr}(T_{\rm Mott,M})}$ ,
while in the former case the sigma meson lies in the 
scattering continuum at all temperatures.
%
%
For the thermodynamics of mesonic correlations in quark matter
we are interested not only in the bound state spectrum of the 
model, but also in the role scattering state continuum. 
This can be consistently discussed with the phase shifts $\phi_M$
parametrizing the complex meson propagator function
\begin{equation}
D_M(z=\omega+i\epsilon,{q})=|D_M(z,{q})|\exp \left[i \phi_M(s)\right]~.
\end{equation}
Since the meson polarization function (\ref{pol})
can be decomposed as
$\Pi_{M}(z,q)=\Pi^{(0)}+a_M(s)\Pi^{(2)}(z,q)$ 
with $a_\pi=s$ and $a_\sigma=s-4m^2$,
an analytic decomposition of the phase shift 
$\phi_M=\phi_{M,R}+\phi_{\rm cont}$ can be made 
\cite{Zhuang:1994dw,Wergieluk:2012gd,Blaschke:2013zaa}.
The continuum phase shift is state independent
%
\begin{equation}
\phi_{\rm cont}(s)=
-\arctan\left[{{\rm Im} \Pi^{(2)}(\omega+i\epsilon,q)}/
{{\rm Re} \Pi^{(2)}(\omega+i\epsilon,q)} \right]~,
\end{equation}
and the resonant phase shift corresponds to a complex pole 
solution $z=z_M=\omega_M +i \Gamma_M/2$ of the Bethe-Salpeter equation
for small width $\Gamma_M$ just above $T_{\rm Mott,M}$, 
which goes over to the bound state pole solution for $\Gamma_M\to 0$,
\begin{equation}
\phi_{M,R}(s)=\pi \Theta(s-M_M^2)~,~~T<T_{\rm Mott,M}~.
\end{equation}
The behaviour of these phase shifts is illustrated in Fig.~\ref{phase0}
for the pion (left panels) and sigma (right panels) meson channels.
These phase shifts obviously obey Levinson's theorem,
\begin{equation}
\int_0^\infty ds \frac{d \phi_M}{ds} = 0 
=  \underbrace{\int_0^{s_{\rm thr}} ds \frac{d \phi_M}{ds}}_{n_M \pi} 
+ \underbrace{\int_{s_{\rm thr}}^\infty ds 
\frac{d \phi_M}{ds}}_{\phi_M(\infty)-\phi_M(s_{\rm thr})}~,
\end{equation}
where $n_M=1$ is the number of bound states below the threshold which for 
the modified NJL model with IR cutoff is 
$s_{\rm thr}(T)=2\sqrt{p^2_{\rm min}+m^2(T)}$.

\begin{figure}[t!hb]
\centerline{
\includegraphics[width=0.9\textwidth,angle=0]{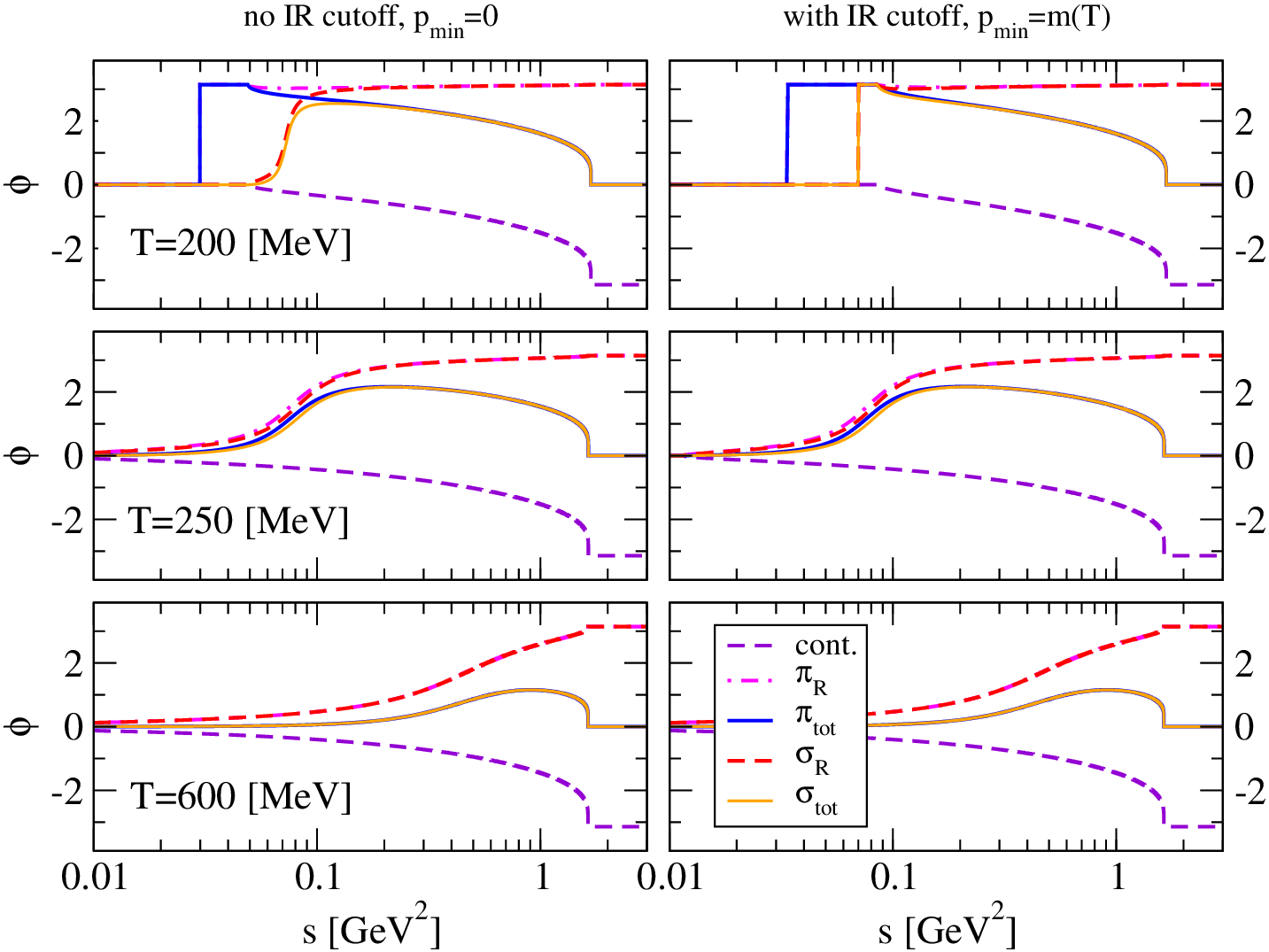}}
\caption{Pion and sigma meson phase shifts for temperatures
200, 250 and 600 MeV with IR cutoff $p_{\rm min}=m(T)$ 
(right panels) and without it (left panels). 
\label{phaseT}
 }
\end{figure}

In Fig.~\ref{phaseT} we show the phase shifts for selected temperatures
around and above the Mott temperature. 
The Levinson theorem holds also in this case and thus the behaviour of the
phase shift at threshold can be used as an indicator for the Mott transition,
i.e., for the transition of a bound state to the scattering state continuum.

\section{Results and Discussion}
We have investigated mesonic correlations in hot quark matter within a 
modified NJL model that incorporates aspects of confinement  by introducing 
a low-momentum cutoff in loop integrals. 
It has the effect to increase the continuum threshold so that both 
chiral partner states, pion and sigma meson are bound states at low 
temperatures. 
Alternatively, this feature is obtained in nonlocal chiral quark models
\cite{Schmidt:1994di,Benic:2013tga}.
Special emphasis in this study was on the Mott dissociation of these states,  
when due to the partial chiral symmetry restoration with 
increasing temperature the threshold of quark-antiquark continuum states drops 
below the meson mass and mesonic correlations change their character from
bound states to resonances at the corresponding Mott temperature.
 
To this end the gap equation for the dynamical quark mass and the polarization 
functions for quark-antiquark correlations have been solved as a function of 
the temperature. 
Hereby we have made the ansatz that the infrared cutoff is identified with 
the dynamically generated quark mass gap thus relating this confinement aspect
with that of chiral symmetry breaking.

The pion and sigma meson correlations are represented by phase shifts which
are decomposed into a resonant and a continuum part which entailing such a 
decomposition also for 
the generalized Beth-Uhlenbeck equation of state.
The continuum contribution is negative and channel independent. 
The form of the resonant phase shifts for the chiral partner states changes 
as the temperature increases from step function in energy that jumps from zero 
to $\pi$ at the meson mass (for mesons at rest) to smoothened step with a 
width $\Gamma$ above the Mott temperature. There, both phases become 
degenerate and thus resemble an aspect of chiral symmetry. 
At the Mott temperature for a given mesonic channel, when the bound state 
vanishes and instead a resonance in the continuum appears, the energy 
derivative of the resonant phase shift changes from a delta-function to a 
Lorentzian (Breit-Wigner) type and the phase shift itself jumps at the 
continuum threshold from $\pi$ to zero in accordance with the Levinson theorem.

\subsection*{Acknowledgements}
The work of A.D. and D.B. was supported in part by the Polish National 
Science Center (NCN) 
under contract number 2011/02/A/ST2/00306 and by the Russian Fund for 
Basic Research under grants 11-02-01538-a (D.B.) 
as well as 12-01-00396 and 13-01-00060 (Yu.L.K.).

\end{document}